\documentclass[manuscript]{aastex631}

\submitjournal{Plan. Sci. J.}

\shorttitle{SEP Track Accumulation in EKB Grains}
\shortauthors{} 

\begin{document}

\title{Solar Energetic Particle Track Accumulation in Edgeworth-Kuiper Belt Dust Grains}

\correspondingauthor{A. R. Poppe}
\email{poppe@berkeley.edu}

\author{M. Lin}
\affiliation{Dept. of Physics, Univ. California at Berkeley, Berkeley, CA, USA}

\author[0000-0001-8137-8176]{A. R. {Poppe}}
\affiliation{Space Sciences Laboratory, Univ. California at Berkeley, Berkeley, CA, USA}



\begin{abstract}

Interplanetary dust grains (IDPs) originate from a variety of sources and are dynamically transported across the solar system.
While in transport, high-$Z$ solar energetic particles (SEPs) with energies of $\sim$1 MeV/nuc leave damage tracks as they pass through IDPs.
SEP track densities can be used as a measure of a grain's space exposure and in turn, help to constrain their lifetimes and origins.
Stratospherically collected IDPs with relatively high track densities ($>10^{10}$ cm$^{-2}$) have been interpreted as originating from the Edgeworth-Kuiper Belt.
To further test this hypothesis, we use a dynamical dust grain tracing model to explore the accumulation of SEP tracks within EKB dust grains.
We demonstrate that, neglecting collisions, dust grains with radii up to 500 $\mu$m are capable of transiting from the EKB to 1 au despite gravitational perturbations from the outer planets, albeit with decreasing probability as a function of size. 
Despite this, we find that EKB grains cannot accumulate sufficient tracks to match those reported in the terrestrial stratospheric IDP collection when applying SEP track accumulation rates established from lunar samples at 1 au and assuming the SEP flux scales with heliocentric distance as $r^{-1.7}$.
By exploring the radial scaling of the SEP flux, we find that a shallower SEP radial distribution of $r^{-1.0}$ does allow for the accumulation of $>$$10^{10}$ tracks cm$^{-2}$ in EKB dust grains that reach 1 au.
We urge further research into the propagation and distribution of high-$Z$ SEPs throughout the heliosphere in order to better constrain track accumulation in IDPs.

\end{abstract}



\section{Introduction} \label{sec:intro}

Interplanetary dust grains (IDPs) originate from a wide variety of sources in the solar system, including asteroids, comets, and Edgeworth-Kuiper Belt (EKB) objects \citep[e.g., see review by][]{Koschny_2019}.
After their birth, these grains are subject to several forces, including gravity, solar radiation pressure, Poynting-Robertson and solar-wind drag, and electromagnetism \citep[e.g.,][]{Burns_1979,Gustafson_1994}, which in turn allows IDPs to migrate long distances from their birthplace \citep[e.g.,][]{Grun_1985, Jackson_1992, Liou_1995, Landgraf_2002,Moro-Martin_2003}.
Thus, the influx of interplanetary dust grains to any object in space is dependent both on the location of such object and the time-integrated dynamics of IDPs originating from various sources.
At Earth, for example, multiple lines of evidence suggest that the IDP influx is primarily composed of dust originating from Jupiter-family comets (JFCs) \citep[e.g.,][]{Nesvorny_2010, Carrillo-Sanchez_2016} with smaller contributions from Oort Cloud comets \citep{Nesvorny_2011}, Halley-type comets \citep{Pokorny_2014}, and the main asteroid belt \citep[e.g.,][]{Durda_1997, Nesvorny_2010}.
In contrast, the IDP influx in the outer solar system is believed to transition from JFC-dominated near Jupiter to EKB-dominated near Neptune \citep[e.g.,][]{Kuchner_2010,Vitense_2012,Poppe_2016, Poppe_2019b, Piquette_2019, Bernardoni_2022}.
Additionally, dynamical simulations have also suggested the possibility that IDPs from the EKB could potentially migrate all the way to 1 au and comprise a component of the zodiacal cloud at Earth \citep[e.g.,][]{Liou_1996, Moro-Martin_2003}; however, supporting evidence for an influx of EKB grains to Earth has, until recently, remained lacking.

As interplanetary dust grains migrate through the solar system, they are subjected to charged-particle irradiation over a wide range of species and energies.
Among these various populations, high-$Z$ ($Z\ge26$) ions with energies of $\sim$1 MeV/nuc leave visible damage tracks within meteoritic minerals due to local melting and subsequent amorphization \citep[e.g.,][]{Fleischer_1975, Szenes_2010}.
These track-inducing particles, comprised mainly of iron (Fe, $Z=26$) with minor contributions from even heavier elements (e.g., Ni, Zn, etc.), primarily originate from solar flares near the Sun's surface or from coronal mass ejections that propagate through the heliosphere and are part of a broader population of `solar energetic particles' (SEPs) \citep[e.g.,][]{Reames_2013, Reames_2019}.
The convolution of the track-inducing SEP distribution throughout the heliosphere and the time history of IDP trajectories governs the rate at which IDPs accumulate damage tracks.
Additionally, energetic particle irradiation of interplanetary dust grains by both SEPs and galactic cosmic rays (GCRs) can also stimulate the formation of cosmogenic radionuclides (e.g., $^{10}$Be, $^{26}$Al, $^{53}$Mn, $^{60}$Fe), observations of which have recently been used to infer an outer solar system origin for several terrestrial micrometeorites \citep[e.g.,][]{feige2024dust}.
While highly complementary to the analysis of SEP-induced track densities studied here, we note that the production of cosmogenic nuclides within IDPs is not the focus of the current study and is left to future work.

%

At Earth, interplanetary dust grains that have entered the terrestrial atmosphere can be collected in the stratosphere for inspection and analysis in ground-based laboratories.
\citet{Bradley_1984} reported the first discovery of SEP-induced tracks in stratospherically collected IDPs and subsequent studies have used SEP track densities to establish the exposure ages of IDPs and by extension, constrain their parent populations \citep[e.g.,][]{Sandford_1986, Flynn_1989, Thiel_1991}.
%
Building on this idea, \cite{flynn1994ekbdust} and \citet{flynn1996ekbdust} suggested that one could identify EKB grains that managed to transit to 1 au based on an unusually high amount of space exposure in the form of SEP-induced tracks.
Recently, \cite{Keller_2022} have reported the possible discovery of such EKB grains in the stratospheric IDP collection, based on a combination of unusually high SEP-induced track densities within some IDPs, $\sim$0.4--5$\times10^{11}$ tracks cm$^{-2}$, and a revised calibration for the SEP track-production rate at 1 au based on lunar samples returned from the Apollo missions \citep{Keller_2021}.
By employing an analytical model of interplanetary dust grain dynamics under the influence of only Poynting-Robertson drag (i.e., without planetary perturbations), \citet{Keller_2022} noted that grains released from the EKB region ($\sim$40-50 au) tended to accumulate between 0.4--4.0$\times10^{10}$ tracks cm$^{-2}$, somewhat lower than their observed distribution.
\citet{Keller_2022} further suggested that, based on the dynamical results of \citet{Liou_1996}, trapping within MMRs outside the orbit of Neptune could yield the additional fraction of accumulated track densities needed to achieve a more satisfactory comparison with the observed high-track-density SEPs; however, an explicit dynamical calculation of this potential effect was not performed.

Here, we build upon and extend the work of \citet{Keller_2022} by using a dynamical dust grain model that includes all major perturbing forces to investigate the dynamics and track accumulation of interplanetary dust grains originating from the Edgeworth-Kuiper Belt.
In particular, we explicitly assess the role that planetary perturbations, including trapping within MMRs, play in the overall accumulation of SEP-induced tracks within IDPs.
Section \ref{sec:model_desc} describes the modeling approach used in the study, including both the interplanetary dust dynamics model and the distribution of track-inducing SEPs.
Section \ref{sec:results} presents the results of our simulations while Section \ref{sec:discuss} provides a discussion of our findings in the context of explaining high-track-density IDPs at 1 au.
Finally, we conclude in Section \ref{sec:conclude}.

\section{Model Description} \label{sec:model_desc}

In order to model the dynamics and SEP-induced track accumulation within interplanetary dust grains, we combine the results from the IDP dynamical model previously described in \citet{Poppe_2016} and \citet{Poppe_2019b} and an analytical description for the rate of SEP track accumulation as a function of heliocentric distance.
Below, we describe both model components in detail.

\subsection{Interplanetary Dust Dynamics Model}

The IDP model uses a dynamic approach to integrate dust-grain state vectors according to the equations of motion with a Bulirsch-Stoer integrator \citep{Press_2007}.
In addition to solar gravity, the dynamical model includes the gravitational forces of the eight planets, solar-radiation pressure, Poynting-Robertson drag, solar wind drag, and the electromagnetic Lorentz force \citep[see full description in][]{Poppe_2016}.
With a specified initial condition (discussed below) and under the combined action of these forces, each individual dust grain is integrated forwards in time until either (i) the dust grain is ejected from the solar system or (ii) the dust grain reaches $<0.05$ au.
During integration, the dust grain state vector is periodically printed to a file for subsequent analysis.
Individual dust grains are initialized with orbital elements drawn from the distributions of EKB objects constrained by the debiased observations reported in \citet{Petit_2011}.
As described in further detail in \citet{Poppe_2016} (see their Appendix A), the full population of EKB objects are separated into four primary sub-populations: classical, scattered, resonant, and outer.
%
%
The relative contributions of each of these EKB sub-populations to the total EKB dust production rates are estimated as: classical, 11\%; scattered, 60\%; resonant, 14\%; and, outer, 15\%.
For this study, we used the library of previously generated EKB interplanetary dust grain state vectors analyzed in the earlier work of \citet{Poppe_2016} and \citet{Poppe_2019b}, which contains a total of approximately 50,000 individual EKB dust grains across 15 discrete, logarithmically spaced grain radii ranging from 2 $\mu$m to 500 $\mu$m.
Dust grains are subjected to erosive mass loss via solar wind sputtering and thus, their radii shrink over time as they transit through interplanetary space. However, this effects tends to only reduce dust grain radii by $\sim$20\% over their lifetimes and thus, does not significantly affect the total accumulation of SEP tracks.
The dust grain material density is assumed to be 2.5 g cm$^{-3}$ consistent with `astrosilicate'-type material.
We neglect the role that grain-grain collisions play on altering the dynamics and lifetimes of IDP grains \citep[e.g.,][]{Stark_2009, Kuchner_2010, Poppe_2016}, leaving this more detailed investigation to future work.

\subsection{SEP Track Accumulation Model}

SEPs are comprised of ions and electrons between energies of ${\sim}$tens of keV to several GeV that originate from solar and heliospheric processes and stream through the heliosphere guided by interplanetary electromagnetic fields \citep[e.g.,][]{Reames_2013, reames2021solar, Klein_2017}. 
SEPs are typically produced via one of two processes: (i) impulsive events that accelerate particles close to the Sun via flares and/or jets, or (ii) gradual events that accelerate particles as coronal mass ejections plow through the corona and interplanetary space \citep{Reames_2013}. 
Compositionally, ion SEPs are dominated by protons, yet also contain species up to Fe ($Z=26$) and beyond \citep{Meyer_etal_1985, Reames_2019}. 
After their initial acceleration, SEPs propagate throughout the heliosphere, guided to first order by the interplanetary Parker spiral magnetic field, yet also subject to other electromagnetic drifts and scattering from turbulence that introduce additional complexity to their overall dynamics \citep{Pei_etal_2006, Kelly_etal_2012, Dalla_etal_2013, Marsh_etal_2013, Laitinen_2016}. 
A key open question in the study of SEP dynamics is the average heliocentric radial scaling of SEP fluxes as a function of ion composition, energy, and solar cycle.
%
Previous work has often focused on the radial behavior of the \textit{peak} ion flux in SEP events \citep[e.g.,][]{lario2007radial, lario2013longitudinal,verkhoglyadova2012radial, He_2017, He_2019} due to its importance in space weather applications, with only more limited investigations into the propagation of high-$Z$ species \citep[e.g.,][]{tylka2013sep,Marsh_etal_2013,dalla2017charge,dalla2017feo}.

In order to model the total accumulation of SEP tracks in IDPs grains, one must adopt a description for the radial dependence of the high-$Z$ SEP flux; however, such a description is not currently available in the literature.
Previous work by \citet{Keller_2022} has thus used the radial scaling of the peak proton intensity as an approximation for the behavior of high-$Z$ SEPs.
%
In particular, they adopted an equation of the form, $R=R_0r^{-\alpha}$, where $R_0$ is the track accumulation rate at 1 au and $\alpha$ is the radial decay exponent.
Based on studies of Apollo sample 64455, \citet{Keller_2021} determined the rate coefficient at 1 au to be $R_o$ = 4.4$\pm$0.4$\times10^4$ cm$^{-2}$ yr$^{-1}$ for a 2$\pi$ exposure (or equivalently, 8.8$\times10^4$ cm$^{-2}$ yr$^{-1}$ for a full 4$\pi$ exposure).
Furthermore, they adopted a value of $\alpha = 1.7$ based on the peak proton flux behavior reported by \citet{He_2017}.
Here, we adopt the same functional form for the SEP accumulation rate as \citet{Keller_2022} in order to first provide a direct comparison; however, as described later in the manuscript, we test the effects of varying the radial exponent $\alpha$ on IDP track accumulation rates. 
We do note that recent analysis of in-situ Fe-group SEPs at 1 au \citep{poppe2023seps} has suggested a higher flux of track-inducing SEPs than found by \citet{Keller_2021} (i.e., the $R_o$ coefficient); however, a reconciliation between these two differing conclusions is not yet in hand. 
Thus, we use the rate from \citet{Keller_2021} in this study to maintain a direct comparison.

Using the dust grain state vectors from the IDP model described above, we derived each grain's heliocentric distance and the corresponding instantaneous track accumulation rate at each time step. 
Integrating over this track accumulation rate for each dust grain thus yields the total tracks accumulated as a function of time.
For this study, the prime periods of interest in each grain's lifetime occur whenever its trajectory crosses Earth's orbit and potentially enters the atmosphere. 
Given the Earth's orbital eccentricity, we selected times when the heliocentric distance of the dust grain lay between 0.982 AU and 1.02 AU (i.e., Earth's perihelion and aphelion) as an instance of the dust grain potentially entering the Earth's atmosphere. 
This method ensures that we retain the rough likelihood of each dust grain entering the Earth as the more instances the dust grain crosses said interval, the higher its chances of entering the Earth's atmosphere.
With this selection, we identified all instances and found the total accumulated track density of the dust grain at each instance. 
In the present exercise, we did not consider the effects of gravitational focusing by the Earth on the relative flux of dust grains entering the atmosphere. 
Such an effect would bias the IDP collection to the grain population with slower average velocities \citep[e.g.,][]{Kortenkamp_2013}; however, such comparative studies are left for future work.

\section{Model Results} \label{sec:results}

\subsection{EKB Transit Probabilities to 1 au}

\begin{figure}[htb]
	\includegraphics*[width=0.7\textwidth]{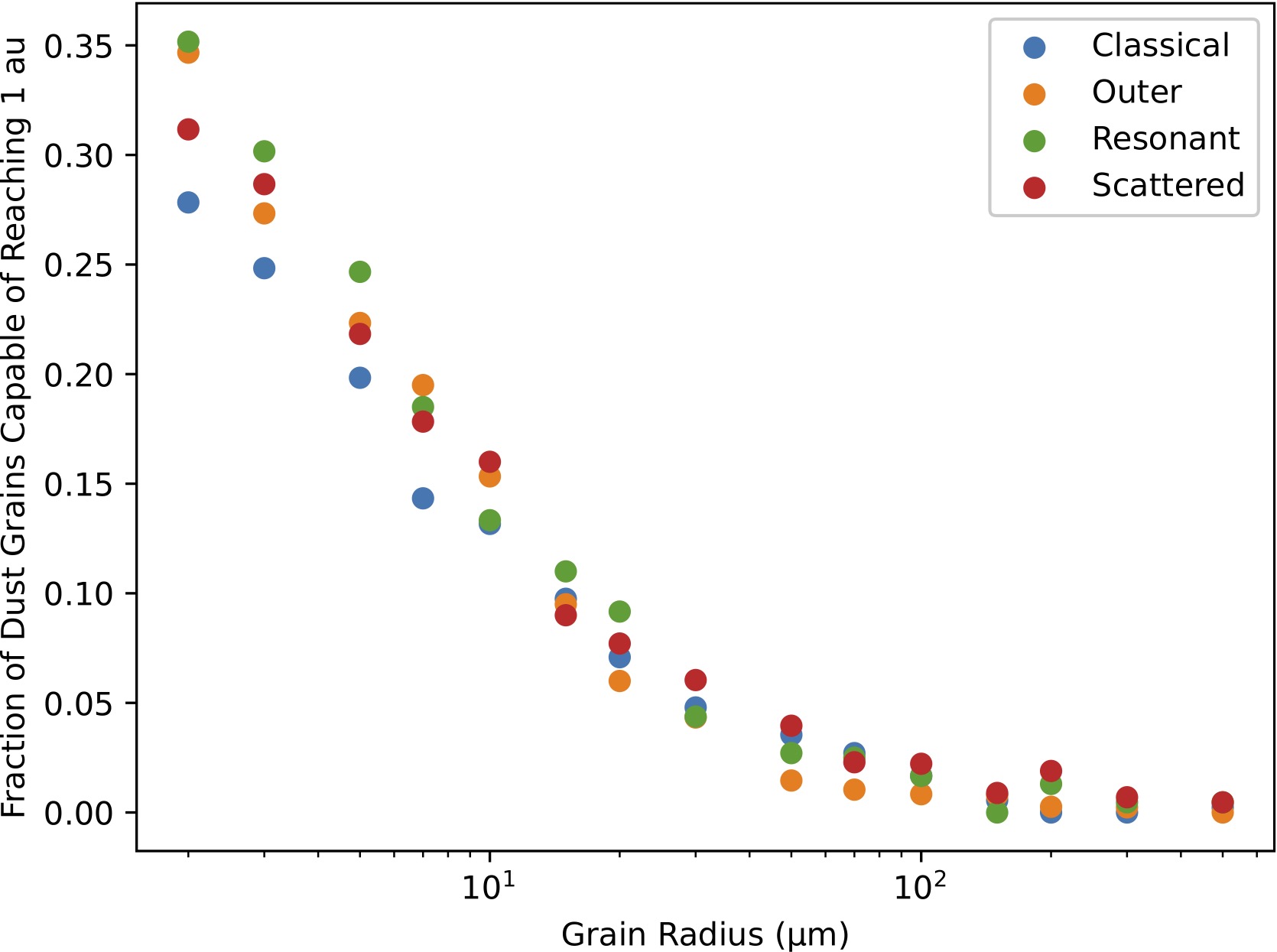}
	\caption{The fraction of modeled IDP grains that migrated from their birthplace in the EKB to 1 au as a function of grain radius for each separate EKB sub-population.}
	\label{fig:fractions}
\end{figure}

Before implementing the SEP track accumulation model, we first analyzed the library of IDP trajectories to determine if, and with what frequency, IDPs originating from the Edgeworth-Kuiper Belt reached 1 au.
As shown in Figure \ref{fig:fractions}, IDPs across all sizes ranges modeled (i.e., 2 $-$ 500 $\mu$m) are capable of reaching 1 au, although Figure \ref{fig:fractions} also shows that the fraction of dust grains that migrated to 1 au decreases as the grain size increases. 
In particular, at 2 $\mu$m, $\sim$30\% of all grains reach 1 au with slight variations seen for different EKB sub-populations. 
For 10 $\mu$m grains, this fraction drops to $\sim$15\%, while grains with radii of 100 $\mu$m and larger only reach 1 au approximately 1-2\% of the time.
%
The overall trend seen in Figure \ref{fig:fractions} is to be expected as Poynting-Robertson drag induces faster inwards drift for smaller dust grains, causing larger dust grains to be comparatively more susceptible to gravitational ejection from the solar system by outer planets, in particular, Neptune \citep[e.g.,][]{Liou_1996, Moro-Martin_2002, Moro-Martin_2003, Poppe_2016}. 
For grain sizes 7 $\mu$m and smaller, we note a trend for fewer classical grains (blue dots) reaching 1 au. This may be caused by the relatively low inclinations and eccentricities of these grains, making them more likely to be captured in mean-motion resonances with Neptune and subsequently ejected from the solar system. 
By the same logic, scattered, resonant, and outer family dust grains may be more likely to reach Earth's orbit due to the opposite reason, i.e., relatively higher eccentricities and inclinations and thus, less susceptibility to trapping within MMRs.
Overall, the model results clearly demonstrate that from a dynamical perspective, IDP grains can successfully migrate from the Edgeworth-Kuiper Belt, past the outer planets, and into the inner solar system with significant fractions.
We note that our results are in agreement with earlier conclusions reached by \citet{Liou_1996} who found that between 15\% and 25\% of EKB grains with diameters between 1 and 9 $\mu$m were able to transit past the outer planets and into 1 au.
Curiously, the results of \citet{Liou_1996} show an \textit{increasing} fraction of grains able to transit to 1 au as a function of diameter as opposed to the \textit{decreasing} trend we find in our results; however, this may be due to the relatively low number of grains simulated in \citet{Liou_1996} (20 grains at each size, compared to $\sim$2500 grains per size in our simulations).

\subsection{SEP Track Accumulation}

\begin{figure}
    \includegraphics*[width=0.85\textwidth]{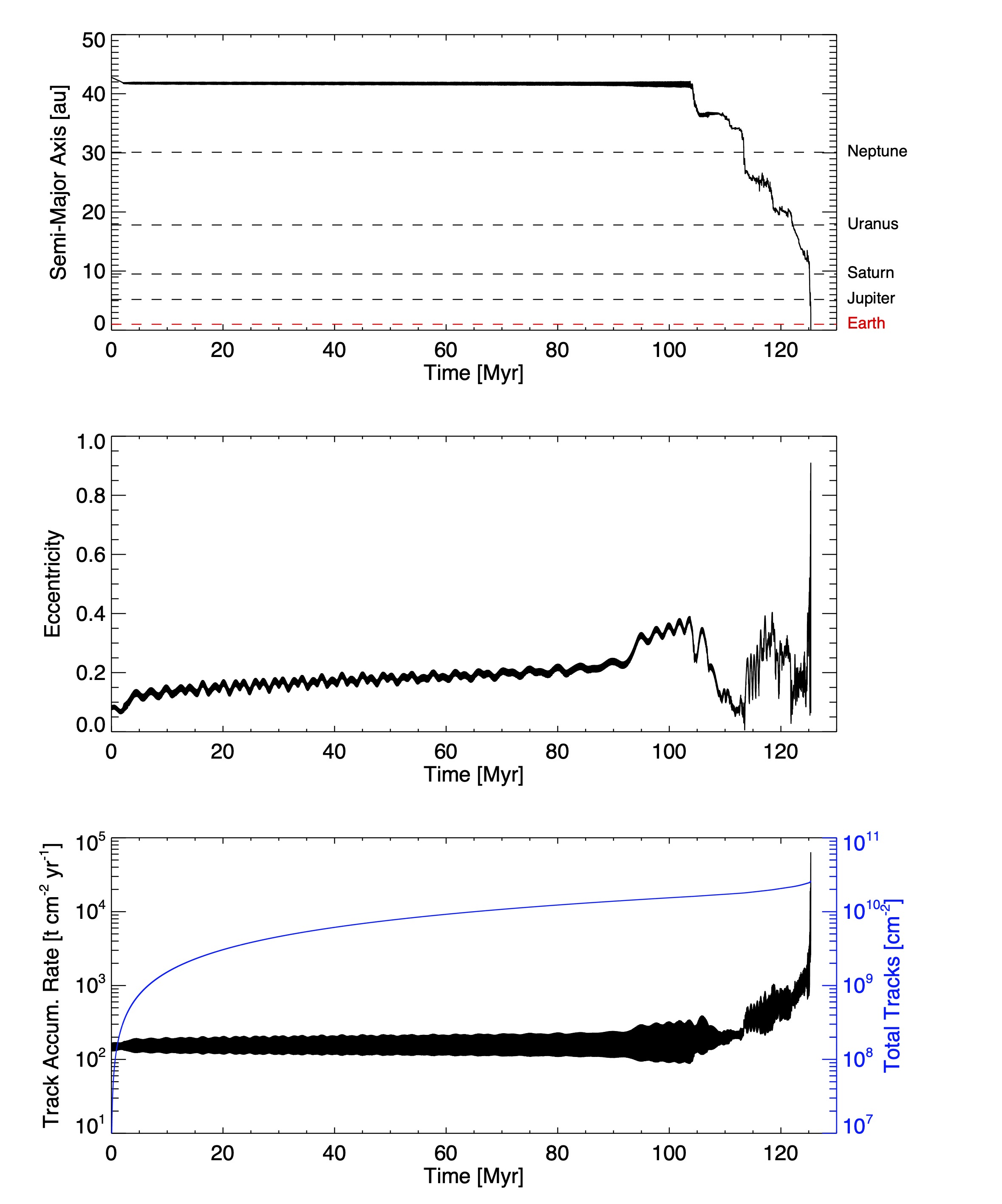}
    \caption{The semi-major axis, eccentricity, track accumulation rate, and total accumulated tracks for a typical 20 $\mu$m EKB classical grain that successfully difted from the EKB to 1 au.}
    \label{fig:example_grain}
\end{figure}

Having established that IDPs can dynamically migrate from the EKB to 1 au, we can now select such grains and analytically model their SEP track accumulation.
Figure \ref{fig:example_grain} shows an example set of results from the IDP and SEP track accumulation models for a 20 $\mu$m EKB-classical grain. 
This particular dust grain was born with a semi-major axis of 42.7 au and an eccentricity of 0.073 and remained within the Kuiper belt for the majority of its lifetime trapped within an MMR with Neptune.
During this trapped period, the grain accumulated tracks at a steady rate of $\sim$150 tracks cm$^{-2}$ yr$^{-1}$.
Just after 100 Myr, it broke free from the mean-motion resonance with Neptune and drifted inwards through the outer solar system over the next $\sim$20 Myr.
During this portion of the grain's lifetime, the track accumulation rate increased up to $\sim$$10^3$ tracks cm$^{-2}$ yr$^{-1}$ as it approached the orbits of Saturn and Jupiter. 
Finally, this grain rapidly entered the inner solar system (i.e., $<$5.2 au) near $\sim$125 Myr before spiraling past the Earth's orbit and into 0.05 au, within only 0.14 Myr.
During this last phase of the grain's lifetime, the track accumulation rate correspondingly increased up to the $8.8\times10^4$ tracks cm$^{-2}$ yr$^{-1}$ at 1 au.
Inspecting the total cumulative tracks gained (blue curve, bottom panel), we see that this grain accumulated a majority of its total tracks during its residence outside of Neptune.
Indeed, 68.6\% of the total tracks accumulated over this grain's lifetime came at distances greater than Neptune (i.e., $>$30 au) while 96.1\% of the total tracks came before the grain passed Jupiter ($>$5.2 au). 
Despite the rapid increase in the track accumulation rate in the inner solar system, the grain only spent $\sim$0.1\% of its lifetime within Jupiter's orbit.
The behavior seen in this example dust grain is in fact fairly typical in our results for those grains that reach 1 au. 
Overall, the majority of tracks accumulated for dust grains in the dynamical model occurred in the outer solar system. 
We also find that this effect is more prominent for larger dust grains since larger fractions of their lifetimes are spent in the outer solar system, typically trapped within MMRs.

\begin{figure}[htb]
	\includegraphics*[width=0.65\textwidth]{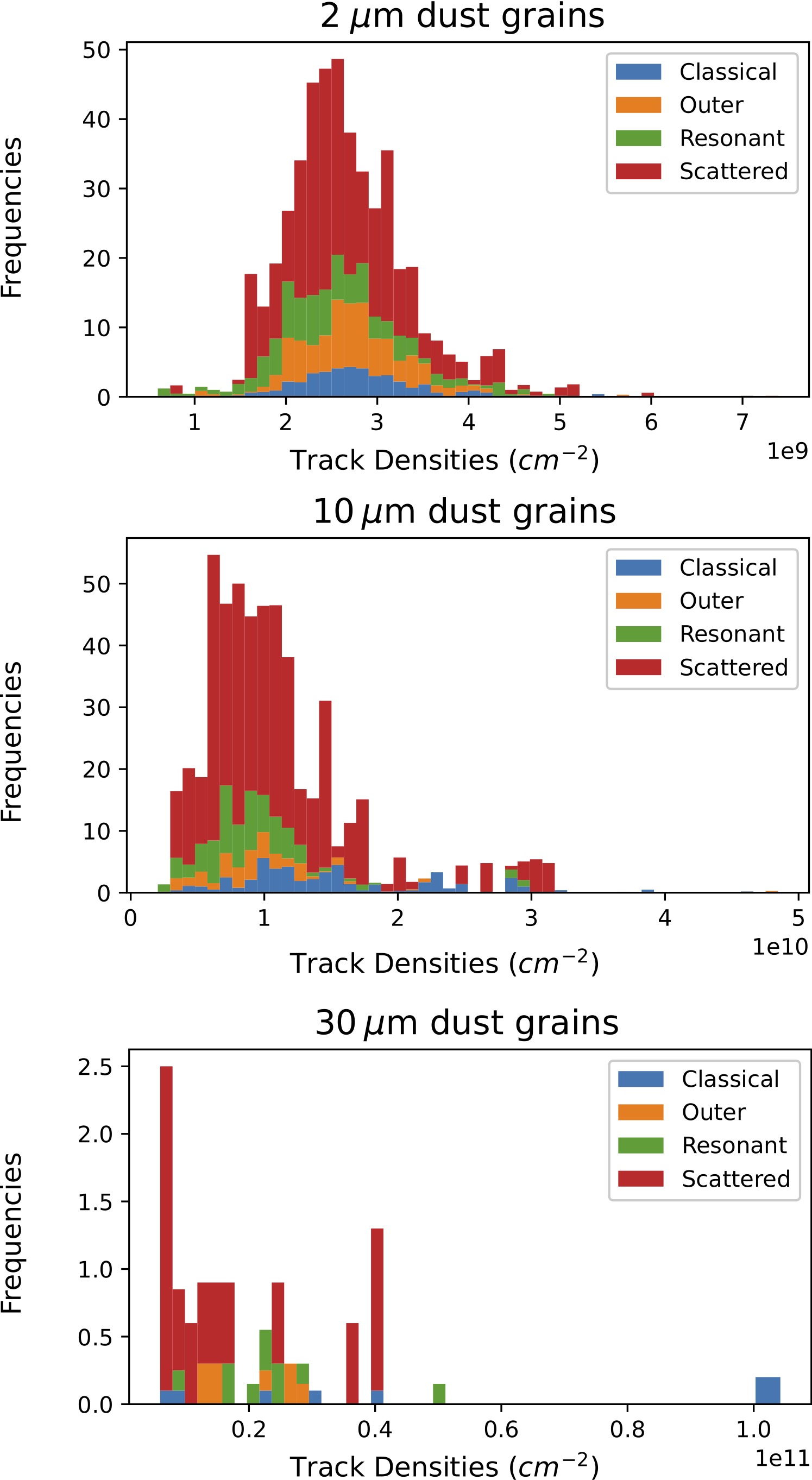}
	\caption{The distributions of dynamically calculated track densities at 1 au for 2, 10, and 30 $\mu$m IDPs. Colors denote the contribution from each EKB sub-population. Note that the distribution of the dust grains is weighted by the relative contribution of each sub-population \citep[see][]{Poppe_2016}.
 }
	\label{fig:track_distributions}
\end{figure}

From a statistical point of view, Figure \ref{fig:track_distributions} shows the distributions of accumulated track densities for those grains that reached 1 au for four separate IDP grain sizes, broken down by EKBO parent family in color. 
Overall, the track densities of dust grains generally increase as a function of dust grain size.
For the 2 $\mu$m sized dust grains, the total tracks accumulated generally ranged from $\sim$1$\times$10$^9$ to $\sim$5$\times$10$^9$ tracks cm$^{-2}$ with a median value of $\sim$2.5$\times$10$^9$ tracks cm$^{-2}$. 
In comparison, the total tracks accumulated in the 10 $\mu$m sized dust grains ranged from $\sim$2$\times$10$^9$ to $\sim$3$\times$10$^{10}$ tracks cm$^{-2}$ with a median of $\sim$1$\times$10$^{10}$ tracks cm$^{-2}$. 
At 30 $\mu$m, where the statistics are noticeably poorer, most track densities are within $\sim$1-4$\times10^{10}$ tracks cm$^{-2}$ with a single outlier near $10^{11}$ tracks cm$^{-2}$.
Indeed, in all four distributions there exist dust grains that accumulated noticeably more tracks than the overall distribution. 
The existence of these outliers is consistent with the observational data of \citet{Keller_2022} that show a small number of grains with comparatively higher track densities.
Despite these outliers, however, the main distributions of track densities in the 30 $\mu$m grain population are lower by a factor of $\sim$2-3 than the observed distribution from \citet{Keller_2022}, which demonstrated a strong clustering near $\sim$6$\times$10$^{10}$ tracks cm$^{-2}$. 
For comparison, the highest track density achieved by any grain that reached 1 au in our model was a 70 $\mu$m radius grain that accumulated $1.8\times10^{11}$ tracks cm$^{-2}$ after an exceptionally long, $\sim$1 Gyr period trapped within an MMR near 42 au; however, such a grain is unlikely to survive collisional grinding in the EKB over such a timescale \citep[e.g.,][]{Kuchner_2010, Poppe_2016, Koschny_2019}.

\subsection{Comparison to Analytical, PR-Drag Only Simulations}

\begin{figure}[htb]
	\includegraphics*[width=0.4\textwidth]{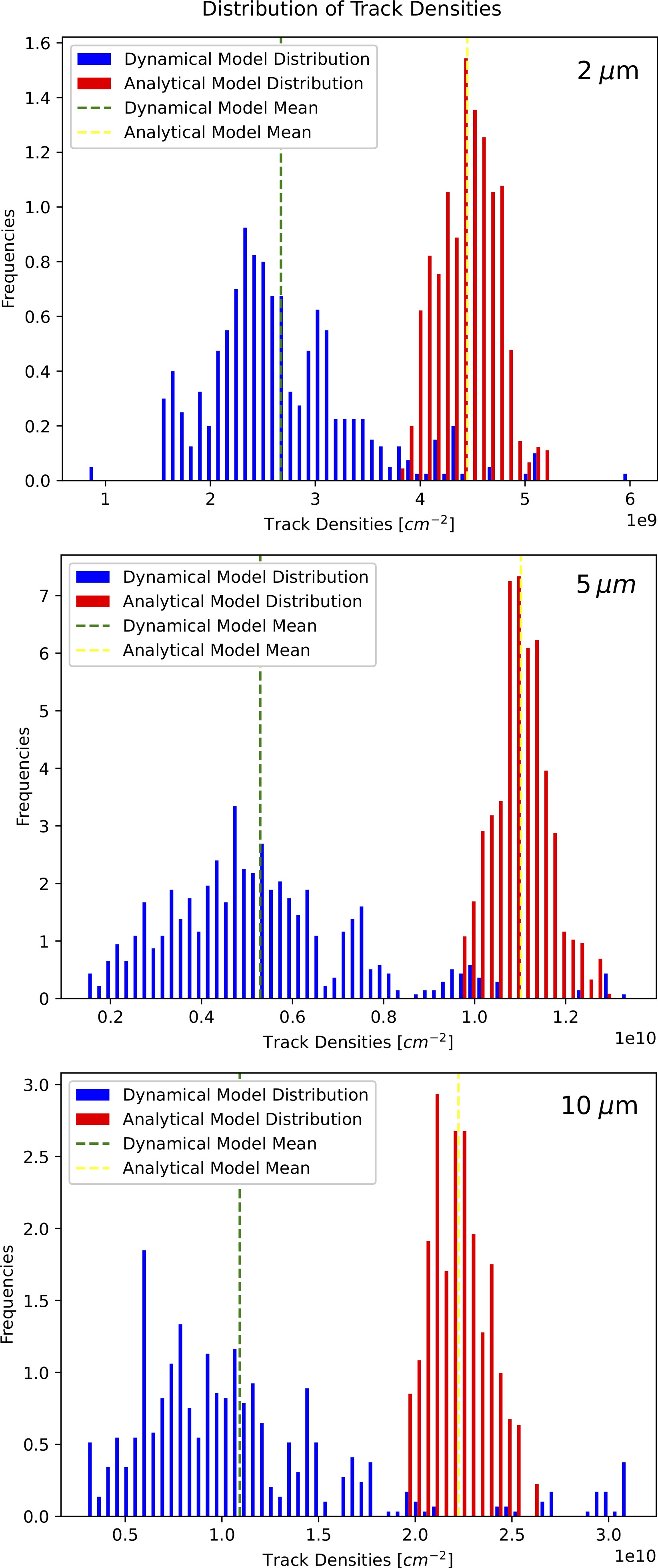}
	\caption{A comparison of the dynamically and analytically calculated track densities in grains reaching 1 au for 2, 5, and 10 $\mu$m IDPs. The vertical dashed lines denote the mean track densities for each distribution.}
	\label{fig:track_dists_comparison}
\end{figure}

\begin{figure}[htb]
	\includegraphics*[width=0.7\textwidth]{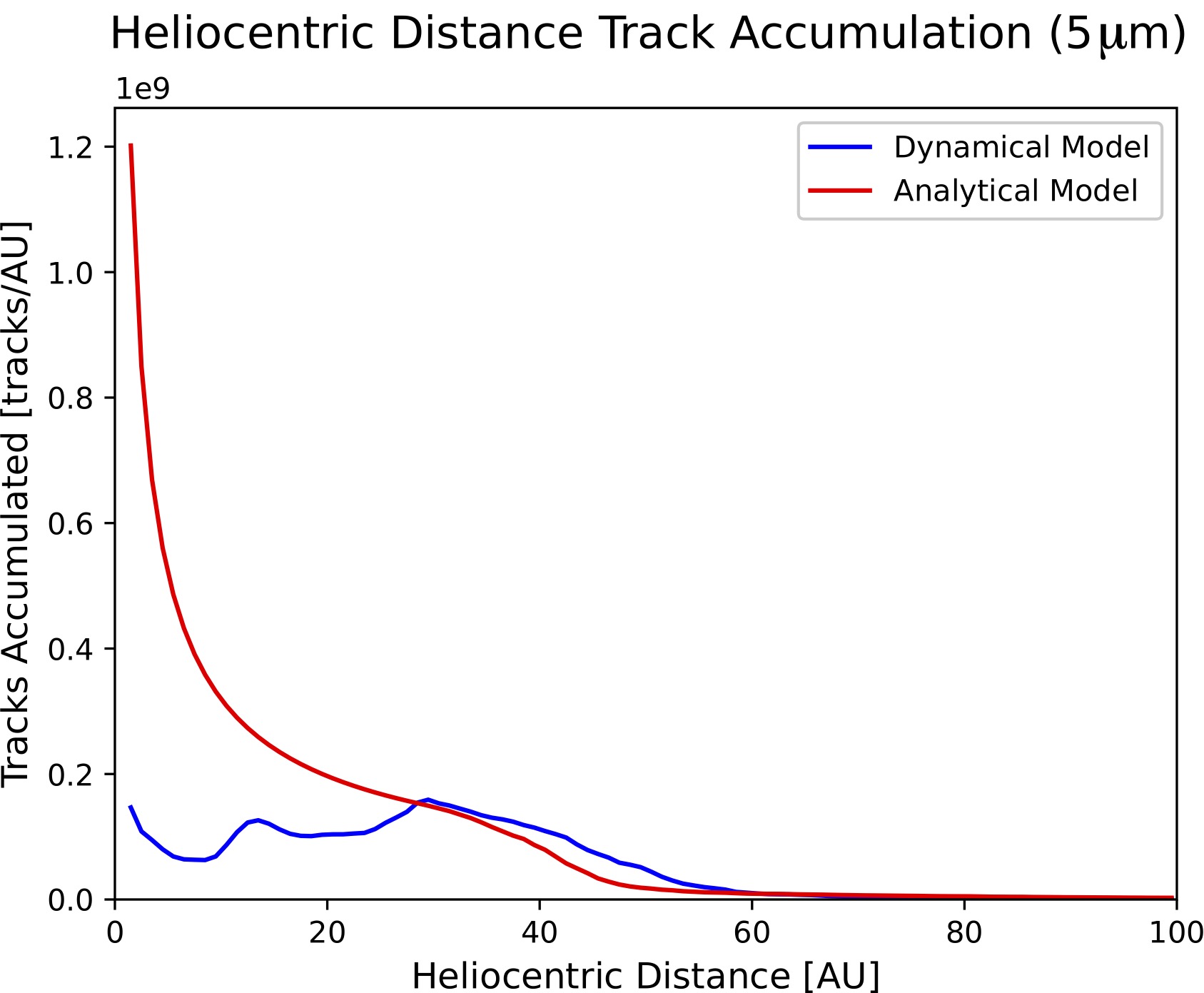}
	\caption{A comparison of the average number of tracks accumulated per au for 5 $\mu$m grains for (blue) dynamical integration and (red) analytical integration.
 }
	\label{fig:avg_tracks_compare}
\end{figure}

To further investigate the apparent discrepancy between our dynamically modeled IDP track density distributions and those measured in the stratospheric dust collection by \citet{Keller_2022}, we conducted an additional set of simulations to calculate track densities under the assumption of only Poynting-Robertson drag (i.e., no planetary perturbations).
This setup, which we term the `analytical model', is similar to the calculations performed in \citet{Keller_2022}, but uses the identical starting orbital conditions for the dynamical model--in other words, for each dust grain we dynamically modeled, we also calculated its trajectory and track accumulation under the assumption of no planetary perturbations, allowing for a 1-to-1 comparison.
%
Figure \ref{fig:track_dists_comparison} shows a comparison of the full dynamical (blue) and analytic (red) track-density distributions for three grain sizes: 2, 5, and 10 $\mu$m grains, respectively. 
Across all three sizes, the track densities of the analytical model are on average greater than that of the dynamical model by a factor of $\sim$2.
We also note that the dynamical model distributions have greater spreads compared to the analytical model, most likely attributable to planetary interactions in the full dynamical model.
To further quantify this, Figure \ref{fig:avg_tracks_compare} compares the average tracks accumulated per au as a function of heliocentric distance for all 5 $\mu$m radius grains between the analytic (red) and dynamic (blue) calculations.
At distances greater than $\sim$30 au, the two curves are similar, with only a slight enhancement in the accumulation of tracks in the dynamical case.
Inwards of 30 au, however, the curves diverge sharply with the dynamical case remaining approximately flat and the analytical case sharply increasing.
Thus, the dynamical interactions with the giant planets slightly increases track accumulation outside of 30 au while strongly suppressing track accumulation inside 30 au.
Because both the dynamical and analytical calculations use the same track accumulation distribution (i.e., $\propto r^{-1.7}$), this difference implies that the dynamically integrated grains are simply spending far less time inside the orbit of Neptune than their analytic counterparts.
Overall, the calculation of the \textit{total} tracks accumulated (taken by the integrals under the two curves, respectively) yields a factor of $\sim$2 difference, in accordance with the distributions shown in Figure \ref{fig:track_dists_comparison}.

\subsection{Variation in SEP Radial Distributions}

\begin{figure}
    \includegraphics*[width=0.7\textwidth]{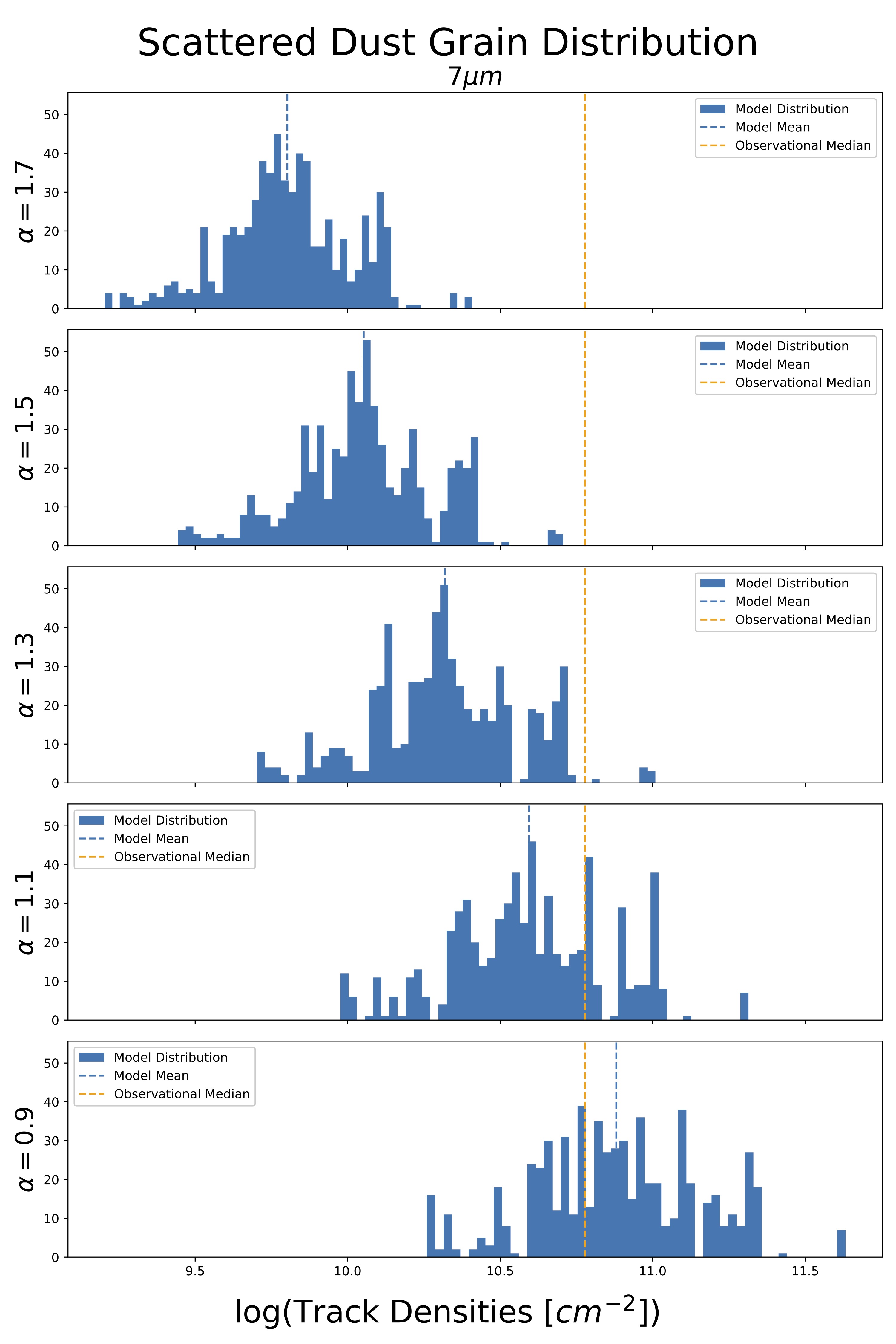}
    \caption{A comparison of the effects of varying the alpha decay constant on the tracks accumulated for 7 $\mu$m size dust grains in the resonant class. The blue vertical dash line shows the average of the track accumulation distribution, while the orange dash line shows the expected mean track density from \cite{Keller_2022}.}
    \label{fig:varying_alpha}
\end{figure}

As discussed in the Introduction, one major uncertainty in the modeling of SEP track accumulation in IDPs is the radial distribution of high-$Z$ SEPs throughout the solar system.
\citet{Keller_2022} adopted a power-law SEP track-production distribution of, $R = R_o r^{-\alpha}$, with $\alpha=1.7$ based on the radial dependence of the peak proton SEP flux \citep{He_2017,He_2019}; however, it is not immediately clear that such choice accurately describes the distribution of high-$Z$ SEPs.
Thus, as a final exercise, we repeated the calculation of accumulated track densities in the subset of EKB grains that reached 1 au over a range of values for the radial exponent, $\alpha$.
In addition to the $\alpha=1.7$ case simulated in our primary results, we also considered exponents of $\alpha = [1.5, 1.3, 1.1, 0.9]$.
The shallower nature of these profiles implies that greater fluxes of high-$Z$ SEPs and the corresponding track-production rate will be increasingly larger at distances $>$1 au (assuming a constant rate coefficient, $R_o$ at 1 au).
Conceptually, the shallower high-$Z$ SEP slope then allows for IDPs to accumulate a greater number of tracks during the portion of their lifetimes that are beyond Neptune.
Figure \ref{fig:varying_alpha} shows the results of this exercise for 7 $\mu$m scattered grains, which are the most abundant dust grain subgroup in the dust grain size range that \cite{Keller_2022} examined.
These results clearly show that as $\alpha$ is progressively lowered, the distribution of accumulated tracks in the 7 $\mu$m scattered grains increases, as expected.
In particular, the results also show that the median modeled track density (blue dashed lines) agrees with the median track density reported by \citet{Keller_2022} between the cases of $\alpha=1.1$ and $\alpha=0.9$.
An analysis of other modeled grain sizes (not shown) yielded similar behavior, namely, an increase in the overall distribution of track densities as a function of decreasing $\alpha$, with improved agreement with the \citet{Keller_2022} observational results approximately between $\alpha=1.1$ and $\alpha=0.9$.

\section{Discussion}
\label{sec:discuss}

The simulations presented above have further quantified both the accessibility of EKB grains to 1 au and the degree of SEP-induced track accumulation such grains gain during their interplanetary transit.
In general agreement with previous simulation studies \citep[e.g.,][]{Liou_1996,Moro-Martin_2003}, we find that EKB grains can transit to 1 au in non-negligible fractions, ranging from $\sim$30\% for 2 $\mu$m grains to $\sim$1-3\% for grains $\simeq$100 $\mu$m when neglecting collisions (although see below for further discussion on this point).
%
%
One question that naturally arises from this finding is the degree to which these EKB grains contribute to the overall mass flux of interplanetary dust to Earth in comparison to that from other sources such as comets (e.g., Jupiter-family, Halley-type, etc.) and asteroids.
If we assume a standard crushing law for the distribution of masses generated in the EKB, and apply both the overall EKB dust mass production rate of $\sim$2$\times10^7$ g s$^{-1}$ reported in \citet{Poppe_2019b} and the relative fraction of grains capable of reaching 1 au as presented in Figure \ref{fig:fractions} here, we estimate a mass accretion rate at Earth of several $\times10^3$ tons yr$^{-1}$.
In comparison, estimates of the total IDP flux to Earth vary from $\sim$$5\times10^4$ tons yr$^{-1}$ from the LDEF experiment \citep{Love_1993}, $\sim$$10^5$ tons yr$^{-1}$ from comparison of a dynamical model to thermal infrared emission \citep{Nesvorny_2010}, to $1.5\times10^4$ tons yr$^{-1}$ as inferred from terrestrial lidar measurements of ablated meteoric material \citep{Carrillo-Sanchez_2016}.  
Thus, we estimate that EKB grains could represent somewhere between 1\% and a few tens of percent to the total mass flux at Earth; however, we generally favor the lower end of this range because, as discussed earlier, this work does not account for grain-grain collisions, which will tend to remove mass from the EKB dust distribution.
More quantitative modeling and analyses of the EKB mass flux contribution to Earth and other inner solar system bodies is clearly warranted in future work.

Despite demonstrating that EKB grains are dynamically capable of transiting to 1 au, our quantitative assessment of track accumulation in EKB grains does not achieve full agreement with the observed track density distributions in stratospheric dust populations reported by \citet{Keller_2022}.
Indeed, a primary finding from our simulations is that the inclusion of planetary perturbations tends to {\it reduce} the accumulated track density in IDPs compared to calculations with only Poynting-Robertson drag.
Our simulations do show that trapping within planetary MMRs, in particular outside Neptune's orbit, does significantly increase the time grains spend at distances $>$30 au; however, track accumulation rates are relatively low at such distances (i.e., assuming an $r^{-1.7}$ scaling, fluxes at 42.5 au are 0.2\% that at 1 au).
Subsequently, our simulations also show that planetary perturbations significantly reduce the amount of time that IDPs originating from the EKB spend transiting through the outer and inner solar systems compared to PR drag, where SEP track accumulation rates are higher.
In total, our dynamically integrated IDPs accumulated a factor of $\sim$2 \textit{less} SEP tracks compared to the analytical, PR-drag only computation. 
Briefly, we note that use of the SEP flux obtained in \citet{poppe2023seps}, which is a factor of $\sim$25$\times$ higher than that determined in \citet{Keller_2021}, would yield model-predicted track density distributions in EKB grains greater than that observed in \citet{Keller_2022}; however, we again note that the nature of the discrepancy in the Fe-group SEP flux at 1 au is not presently understood.
Thus, in light of our finding of fewer tracks in EKB grains at 1 au with the use of the \citet{Keller_2021} track-inducing SEP flux, we consider below several additional ideas to explore in order to bring about a better understanding of the origin(s) of high track density IDPs collected at 1 au.

First, the results of our simulations have demonstrated a critical need for a better understanding of high-$Z$ SEP dynamics throughout the heliosphere.
Due to a lack of such understanding, previous work has relied on observations and modeling of the behavior of the peak proton SEP flux as a function of heliocentric radial distance \citep[e.g.,][]{He_2017,He_2019}; however, it is possible, if not likely, that the average $Z>26$ SEP flux maintains a different radial scaling.
Indeed, full-particle integration of both proton and Fe SEPs through Parker-spiral interplanetary magnetic fields has demonstrated varying degrees of electromagnetic drift as a function of the SEP charge-to-mass ratio \citep{Marsh_etal_2013}.
A full understanding of high-$Z$ SEP dynamics would also account for variations in solar cycle including the effects of focusing versus defocusing interplanetary magnetic fields, the degree of scattering due to interplanetary turbulence, and the distribution of SEP source locations.
Such simulations could be tested against our findings that a relatively shallower high-$Z$ SEP profile with exponential slope, $\alpha$$\sim$$1.0$ (i.e., see Figure \ref{fig:varying_alpha}), allows for a consistent data-model comparison for accumulated SEP tracks in EKB grains.

While high-$Z$ SEPs have been assumed here as the sole source of tracks observed within IDPs, one may consider other sources of high-energy particles that may contribute to track formation.
\citet{Keller_2022} considered the possibility that galactic cosmic rays (GCRs) could contribute to track formation, yet concluded this was unlikely as GCRs must penetrate on the order of meters in depth before depositing their energy and inducing track formation, a far greater depth than IDP radii considered here.
Despite this, it is theoretically possible that GCR irradiation of EKB parent bodies with sizes meters and larger could provide an initial set of GCR-induced tracks before an individual EKB dust grain was generated (e.g., via mutual EKB parent body collisions) and released into interplanetary space.
%
Stochastic increases in the GCR flux in the EKB could also occur during periods where the heliosphere passed through dense and/or cold interstellar clouds \citep[e.g.,][]{Muller_2006, Muller_2008, opher2024heliosphere} or was subjected to supernovae ejecta \citep{Fields_2008,miller2022supernova}.
Either type of event could expose both the EKB parent bodies and the interplanetary dust populations in the outer solar system to increased fluxes of energetic particles; however, the present uncertainty in the timing, duration, and energetic particle intensity of such events currently prevents us from quantitatively assessing their potential contribution to track formation within EKB IDPs.

An additional source not considered by \citet{Keller_2022} that are within the appropriate energy range for generating track formation are anomalous cosmic rays (ACRs).
ACRs are a population of approximately 0.5$-$50 MeV/nuc energetic ions that likely originate from the acceleration of interstellar and interplanetary pickup ions at the distant heliospheric termination shock (or somewhere beyond) that stream inwards into the heliosphere proper \citep[e.g.,][]{klecker1998acrs, jokipii1998acrs, giacalone2022acrs}. 
The seed populations of ACRs come from three primary sources: (i) inflowing interstellar neutrals \citep[e.g.,][]{fisk1974acrs}, (ii) an `inner' source of neutrals originating from solar wind interactions with interplanetary dust \citep[e.g.,][]{gloeckler1998inner,Schwadron_2000,Quinn_2018}, and (iii) an `outer' source of neutrals from solar wind interactions with Edgeworth-Kuiper Belt dust grains \citep[e.g.,][]{Schwadron_2002, Schwadron_2007}.
ACRs are comprised of a broad composition reflecting both the interstellar neutral gas and solar wind sources from which they arise \citep[e.g.,][]{Reames_1999, cummings2007acrcomp}.
ACRs up to Ar ($Z=18$) have been reported at 1 au \citep{Reames_1999} while analysis of Voyager observations have reported ACR detections up to Fe ($Z=26$) \citep{stone1997acrs}.
Despite these Voyager observations, however, little remains known about the full intensity and distribution of Fe ACRs in the outer heliosphere.
A population of Fe ACRs could contribute additional track formation to interplanetary grains, in particular those grains that originate from the EKB outwards, as ACR fluxes increase as one approaches the Termination Shock \citep[located near $\sim$70-150 au;][]{McComas_2019}.
Further measurements and/or modeling of Fe (and heavier) ACRs in the outer solar system are clearly needed to better understand the potential contribution of Fe ACRs on IDP track accumulation rates.

We have neglected grain-grain collisions in our analysis here, however, they are likely to play some role in the distribution of track densities for grains that transit to 1 au.
First, grain-grain collisions may decrease the fraction of EKB grains that successfully transit to 1 au, although previous simulations that implemented collisional-grooming algorithms \citep[e.g.,][]{Poppe_2016} still demonstrate a non-zero flux of EKB IDPs to 1 au.
While not explicitly calculated in \citet{Poppe_2016}, estimates of the EKB mass flux to 1 au in the fully collisional case are on the order of $\sim$1\% the total mass to Earth \citep[e.g.,][]{Nesvorny_2010, Carrillo-Sanchez_2016}. The relatively low value of this estimate is primarily dictated by the collisional destruction of larger dust grains that possess the majority of the IDP mass flux.
Grain-grain collisions may also affect the total accumulation of SEP tracks in IDPs, albeit in two competing directions.
First, grain-grain collisions can effectively limit the lifetimes of grains entrained in MMRs, at least for grains with radii larger than $\sim$5 $\mu$m \citep[i.e., see Figure 10,][]{Koschny_2019}.
Thus, the ``enhancement" in SEP track densities that grains $>$5 $\mu$m may theoretically receive while trapped in MMRs could be severely limited as such grains are collisionally ground down on faster timescales.
In contrast, however, grain-grain collisions may instead offer a path for an accumulation of SEP tracks starting in very large (e.g., $\sim$1 mm) grains that are then successively ground down in a cascade of smaller grains, each of which accumulates tracks during its own lifetime before disruption.
A more quantitative assessment of these processes requires a combination of a collisional-grooming algorithm for interplanetary dust grains and a Monte Carlo-type method for simulating SEP track accumulation in individual grains as they are collisionally broken up.

If the primary populations of the Edgeworth-Kuiper Belt are not the source of high-track-density IDPs, one could consider alternative sources.
\citet{Keller_2022} briefly considered Oort Cloud cometary (OCC) grains as a possible source, yet dismissed it as unlikely due to expectations that OCC grains cross Earth's orbit at relatively high speeds and thus, are likely to suffer thermal annealing effects during atmospheric entry.
Indeed, dynamical simulations of OCC grains at 1 au tend to support this hypothesis \citep[e.g.,][]{Nesvorny_2011, Pokorny_2018} and barring any new simulations that demonstrated a dynamical pathway for OCC grains to arrive at 1 au on relatively low-eccentricity, low-inclination orbits, we would agree with the conclusion that OCC grains are a less likely source high-track-density grains than the EKB.
To be clear, we do note that neither \citet{Keller_2022} nor our work here has actually calculated the predicted SEP track densities within OCC grains that reach 1 au. Such an exercise may be worthwhile to at least establish to first order if OCC grains can accumulate the observed number of tracks. If they do, this may provide motivation to revisit more critically the assumption that high velocity OCC grains will not maintain a full record of their tracks as they enter Earth's atmosphere and decelerate.

Apart from OCC grains, one could also consider the possibility of the existence of a source of dust grains more distant than the main Edgeworth-Kuiper Belt on low-eccentricity, low-inclination orbits.
While the `outer' EKB dust grain population that we have considered here does not accumulate sufficient tracks (i.e., see the orange distribution in Figure \ref{fig:track_distributions}), these outer grains are launched on relatively high eccentricity orbits that tend to bring their perihelia close to Neptune (between $\sim$30-40 au).
These close encounters with Neptune tend to scatter the grains prematurely and thus, limit the degree of tracks they can accumulate.
If, however, there was a low-eccentricity component to the outer EKB, such grains may remain dynamically decoupled from Neptune for much longer periods of time, allowing them to accumulate a greater number of tracks as they slowly drift inwards under the influence of P-R drag.
For example, using the analytic approach to calculating track densities, a 25 $\mu$m radius, 2.0 g cm$^{-3}$ grain would accumulate $\sim$6$\times10^{10}$ tracks cm$^{-2}$ (near the median observed by \citet{Keller_2022}) if it originated at 100 au on a zero-eccentricity orbit.
%
Curiously, we do note that the New Horizons Student Dust Counter, which is currently transiting through the EKB near $\sim$60 au, has recently reported higher than expected dust fluxes compared to earlier model predictions \citep{doner2024sdc}.
While a full understanding of this model-data discrepancy is an active area of research, one hypothesis suggested by \citet{doner2024sdc} is the presence of a more distant population of the EKB as an additional, and previously unaccounted for, source of IDPs \citep[see also][]{fraser2023ekb}.
Continued exploration of the Edgeworth-Kuiper Belt and its ultimate extent will help to further test this intriguing hypothesis.

\section{Conclusion}
\label{sec:conclude}

A study of SEP-induced track accumulation in interplanetary dust grains originating from the Edgeworth-Kuiper Belt has shown that such grains do not accumulate sufficient tracks in comparison to high-track-density grains reported in the stratospheric dust collection at 1 au \citep{Keller_2022}.
In fact, dynamically integrated trajectories for EKB IDPs tend to accumulate a factor of approximately two \textit{less} track density compared to analytical integrations with only Poynting-Robertson drag.
Thus, with our current understanding of outer solar system dust dynamics as well as the magnitude and radial distribution of the flux of high-$Z$ SEPs throughout the heliosphere, the origin of such high-track-density grains remains an open problem.
Despite this, research into several key areas could further improve and/or change our understanding.
In particular, a better understanding of the dynamics of high-$Z$ ($Z>26$) SEPs throughout the heliosphere and the resulting average flux experienced by interplanetary dust grains is of key importance.
Simulations of these SEP dynamics are well within reach using current computational methods \citep[e.g.,][]{Marsh_etal_2013, dalla2017charge, dalla2017feo}.
Additionally, the presence of Fe and heavier anomalous cosmic rays in the outer heliosphere \citep[e.g.,][]{stone1997acrs} may be an additional source of tracks within IDPs. 
Further research into the flux and distribution of these ACR populations is thus warranted.
One could also consider exploring how variations in the assumed IDP particle properties (e.g., material density, optical scattering properties) that control perturbing forces such as Poynting-Robertson and/or solar wind drag affect the overall track accumulation.
Finally, a full reconciliation between lunar-derived SEP-induced track rates \citep{Keller_2021} and in-situ Fe-group SEP flux measurements \citep{poppe2023seps} is also required to fully understand SEP track accumulation in interplanetary dust grains.



\section{Acknowledgements}
The authors acknowledge support from NASA's New Frontiers Data Analysis Program, grant \#80NSSC18K1557.
The authors also thank the reviewers for helpful comments that improved the manuscript.
\bibliographystyle{aasjournal}

\begin{thebibliography}{}
\expandafter\ifx\csname natexlab\endcsname\relax\def\natexlab#1{#1}\fi
\providecommand{\url}[1]{\href{#1}{#1}}
\providecommand{\dodoi}[1]{doi:~\href{http://doi.org/#1}{\nolinkurl{#1}}}
\providecommand{\doeprint}[1]{\href{http://ascl.net/#1}{\nolinkurl{http://ascl.net/#1}}}
\providecommand{\doarXiv}[1]{\href{https://arxiv.org/abs/#1}{\nolinkurl{https://arxiv.org/abs/#1}}}

\bibitem[{Bernardoni {et~al.}(2022)Bernardoni, Hor{\'a}nyi, Doner, Piquette,
  Szalay, Poppe, James, Hunziker, Sterken, Strub, Olkin, Singer, Spencer,
  Stern, \& Weaver}]{Bernardoni_2022}
Bernardoni, E., Hor{\'a}nyi, M., Doner, A., {et~al.} 2022, Plan. Sci. J., 3,
  \dodoi{10.3847/PSJ/ac5ab7}

\bibitem[{Bradley {et~al.}(1984)Bradley, Brownlee, \& Fraundorf}]{Bradley_1984}
Bradley, J.~P., Brownlee, D.~E., \& Fraundorf, P. 1984, Science, 226, 1432,
  \dodoi{10.1126/science.226.4681.1432}

\bibitem[{Burns {et~al.}(1979)Burns, Lamy, \& Soter}]{Burns_1979}
Burns, J.~A., Lamy, P.~L., \& Soter, S. 1979, Icarus, 40, 1

\bibitem[{{Carrillo-S\'anchez} {et~al.}(2016){Carrillo-S\'anchez},
  Nesvorn{\'y}, Pokorn{\'y}, Janches, \& Plane}]{Carrillo-Sanchez_2016}
{Carrillo-S\'anchez}, J.~D., Nesvorn{\'y}, D., Pokorn{\'y}, P., Janches, D., \&
  Plane, J. M.~C. 2016, Geophys. Res. Lett., 43, 11979

\bibitem[{Cummings \& Stone(2007)}]{cummings2007acrcomp}
Cummings, A.~C., \& Stone, E.~C. 2007, Space Sci. Rev., 130, 389,
  \dodoi{10.1007/s11214-007-9161-y}

\bibitem[{Dalla {et~al.}(2017{\natexlab{a}})Dalla, Marsh, \&
  Battarbee}]{dalla2017charge}
Dalla, S., Marsh, M.~S., \& Battarbee, M. 2017{\natexlab{a}}, Astrophys. J.,
  834, \dodoi{10.3847/1538-4357/834/2/167}

\bibitem[{{Dalla} {et~al.}(2013){Dalla}, {Marsh}, {Kelly}, \&
  {Laitinen}}]{Dalla_etal_2013}
{Dalla}, S., {Marsh}, M.~S., {Kelly}, J., \& {Laitinen}, T. 2013, Journal of
  Geophysical Research (Space Physics), 118, 5979, \dodoi{10.1002/jgra.50589}

\bibitem[{Dalla {et~al.}(2017{\natexlab{b}})Dalla, Marsh, Zelina, \&
  Laitinen}]{dalla2017feo}
Dalla, S., Marsh, M.~S., Zelina, P., \& Laitinen, T. 2017{\natexlab{b}},
  Astron. Astrophys., 598, \dodoi{10.1051/0004-6361/201628618}

\bibitem[{Doner {et~al.}(2024)Doner, Hor{\'a}nyi, Bagenal, Brandt, Grundy,
  Lisse, Parker, Poppe, Singer, Stern, \& Verbiscer}]{doner2024sdc}
Doner, A., Hor{\'a}nyi, M., Bagenal, F., {et~al.} 2024, Astrophys. J. Lett.,
  961, \dodoi{10.3847/2041-8213/ad18b0}

\bibitem[{Durda \& Dermott(1997)}]{Durda_1997}
Durda, D.~D., \& Dermott, S.~F. 1997, Icarus, 130, 140,
  \dodoi{https://doi.org/10.1006/icar.1997.5803}

\bibitem[{Feige {et~al.}(2024)Feige, Airo, Berger, Br\"uckner, G\"artner,
  Genge, Leya, {Habibi Marekani}, Hecht, Klingner, Lachner, Li, Merchel,
  Nissen, Patzer, Peterson, Schropp, Sager, Suttle, Trappitsch, \&
  Weinhold}]{feige2024dust}
Feige, J., Airo, A., Berger, D., {et~al.} 2024, Phil. Trans. R. Soc. A, 382,
  \dodoi{10.1098/rsta.2023.0197}

\bibitem[{Fields {et~al.}(2008)Fields, Athanassiadou, \& Johnson}]{Fields_2008}
Fields, B.~D., Athanassiadou, T., \& Johnson, S.~R. 2008, Astrophys. J., 678

\bibitem[{Fisk {et~al.}(1974)Fisk, Kozlovsky, \& Ramaty}]{fisk1974acrs}
Fisk, L.~A., Kozlovsky, B., \& Ramaty, R. 1974, Astrophys. J., 190,
  \dodoi{10.1086/181498}

\bibitem[{Fleischer {et~al.}(1975)Fleischer, Price, \& Walker}]{Fleischer_1975}
Fleischer, R.~L., Price, P.~B., \& Walker, R.~M. 1975, {Nuclear Tracks in
  Solids: Principles and Applications} ({University of California Press})

\bibitem[{Flynn(1989)}]{Flynn_1989}
Flynn, G.~J. 1989, Icarus, 77, 287, \dodoi{10.1016/0019-1035(89)90091-2}

\bibitem[{Flynn(1994)}]{flynn1994ekbdust}
Flynn, G.~J. 1994, in {Abstracts of the 25th Lunar and Planetary Science
  Conference}, Vol.~25

\bibitem[{Flynn(1996)}]{flynn1996ekbdust}
Flynn, G.~J. 1996, in {ASP Conference Series}, Vol. 104, {Physics, Chemistry,
  and Dynamics of Interplanetary Dust}, ed. B.~A.~S. Gustafson \& M.~S. Hanner

\bibitem[{{Fraser} {et~al.}(2023){Fraser}, {Porter}, {Lin}, {Napier},
  {Spencer}, {Kavelaars}, {Verbiscer}, {Yoshida}, {Terai}, {Ito}, {Gerdes},
  {Benecchi}, {Stern}, {Gwyn}, {Buie}, {Peltier}, {Singer}, {Brandy}, {New
  Horizons Lorri Team}, \& {New Horizons Ggi Science Team}}]{fraser2023ekb}
{Fraser}, W.~C., {Porter}, S.~B., {Lin}, H.~W., {et~al.} 2023, in LPI
  Contributions, Vol. 2806, 54th Lunar and Planetary Science Conference, 2361

\bibitem[{Giacalone {et~al.}(2022)Giacalone, Fahr, Fichtner, Heber, Hill,
  K\'ota, Leske, Potgieter, \& Rankin}]{giacalone2022acrs}
Giacalone, J., Fahr, H., Fichtner, H., {et~al.} 2022, Space Sci. Rev., 218,
  \dodoi{10.1007/s11214-022-00890-7}

\bibitem[{Gloeckler \& Geiss(1998)}]{gloeckler1998inner}
Gloeckler, G., \& Geiss, J. 1998, Space Sci. Rev., 86, 127,
  \dodoi{10.1023/A:1005019628054}

\bibitem[{Gr{\"u}n {et~al.}(1985)Gr{\"u}n, Zook, Fechtig, \& Giese}]{Grun_1985}
Gr{\"u}n, E., Zook, H.~A., Fechtig, H., \& Giese, R.~H. 1985, Icarus, 62, 244

\bibitem[{Gustafson(1994)}]{Gustafson_1994}
Gustafson, B. A.~S. 1994, Annu. Rev. Earth Planet. Sci., 22, 553

\bibitem[{He \& Wan(2019)}]{He_2019}
He, H.-Q., \& Wan, W. 2019, Astrophys. J. Lett., 885,
  \dodoi{10.3847/2041-8213/ab50bd}

\bibitem[{He {et~al.}(2017)He, Zhou, \& Wan}]{He_2017}
He, H.-Q., Zhou, G., \& Wan, W. 2017, Astrophys. J., 842,
  \dodoi{10.3847/1538-4357/aa7574}

\bibitem[{Jackson \& Zook(1992)}]{Jackson_1992}
Jackson, A.~A., \& Zook, H.~A. 1992, Icarus, 97, 70

\bibitem[{Jokipii \& Giacalone(1998)}]{jokipii1998acrs}
Jokipii, J.~R., \& Giacalone, J. 1998, Space Sci. Rev., 83, 123,
  \dodoi{10.1023/A:1005077629875}

\bibitem[{Keller {et~al.}(2021)Keller, Berger, Zhang, \&
  Christoffersen}]{Keller_2021}
Keller, L.~P., Berger, E.~L., Zhang, S., \& Christoffersen, R. 2021, Meteorit.
  Planet. Sci., 56, 1685, \dodoi{10.1111/maps.13732}

\bibitem[{Keller \& Flynn(2022)}]{Keller_2022}
Keller, L.~P., \& Flynn, G.~J. 2022, Nature Astron., 6, 731,
  \dodoi{10.1038/s41550-022-01647-6}

\bibitem[{{Kelly} {et~al.}(2012){Kelly}, {Dalla}, \&
  {Laitinen}}]{Kelly_etal_2012}
{Kelly}, J., {Dalla}, S., \& {Laitinen}, T. 2012, \apj, 750, 47,
  \dodoi{10.1088/0004-637X/750/1/47}

\bibitem[{Klecker {et~al.}(1998)Klecker, Mewaldt, Bieber, Cummings, Drury,
  Giacalone, Jokipii, Jones, Krainev, Lee, {et~al.}}]{klecker1998acrs}
Klecker, B., Mewaldt, R., Bieber, J., {et~al.} 1998, Space Science Reviews, 83,
  259

\bibitem[{Klein \& Dalla(2017)}]{Klein_2017}
Klein, K.-L., \& Dalla, S. 2017, Space Sci. Rev., 212, 1107,
  \dodoi{10.1007/s11214-017-0382-4}

\bibitem[{Kortenkamp(2013)}]{Kortenkamp_2013}
Kortenkamp, S.~J. 2013, Icarus, 226, 1550, \dodoi{10.1016/j.icarus.2013.08.020}

\bibitem[{Koschny {et~al.}(2019)Koschny, Soja, Engrand, Flynn, Lasue,
  Levasseur-Regourd, Malaspina, Nakamura, Poppe, Sterken, \&
  Trigo-Rodriguez}]{Koschny_2019}
Koschny, D., Soja, R.~H., Engrand, C., {et~al.} 2019, Space Sci. Rev., 215,
  \dodoi{10.1007/s11214-019-0597-7}

\bibitem[{Kuchner \& Stark(2010)}]{Kuchner_2010}
Kuchner, M.~J., \& Stark, C.~C. 2010, Astron. J., 140, 1007

\bibitem[{Laitinen {et~al.}(2016)Laitinen, Kopp, Effenberger, Dalla, \&
  Marsh}]{Laitinen_2016}
Laitinen, T., Kopp, A., Effenberger, F., Dalla, S., \& Marsh, M.~S. 2016,
  Astron. Astrophys., 591, \dodoi{10.1051/0004-6361/201527801}

\bibitem[{Landgraf {et~al.}(2002)Landgraf, Liou, Zook, \&
  Gr{\"u}n}]{Landgraf_2002}
Landgraf, M., Liou, J.-C., Zook, H.~A., \& Gr{\"u}n, E. 2002, The Astronomical
  Journal, 123, 2857

\bibitem[{Lario {et~al.}(2007)Lario, Aran, Agueda, \&
  Sanahuja}]{lario2007radial}
Lario, D., Aran, A., Agueda, N., \& Sanahuja, B. 2007, Adv. Space Res., 40,
  289, \dodoi{10.1016/j.asr.2007.01.057}

\bibitem[{Lario {et~al.}(2013)Lario, Aran, {G\'omez-Herrero}, Dresing, Heber,
  Ho, Decker, \& Roelof}]{lario2013longitudinal}
Lario, D., Aran, A., {G\'omez-Herrero}, R., {et~al.} 2013, Astrophys. J., 767,
  \dodoi{10.1088/0004-637X/767/1/41}

\bibitem[{Liou {et~al.}(1996)Liou, Zook, \& Dermott}]{Liou_1996}
Liou, J.-C., Zook, H.~A., \& Dermott, S.~F. 1996, Icarus, 124, 429

\bibitem[{Liou {et~al.}(1995)Liou, Zook, \& Jackson}]{Liou_1995}
Liou, J.-C., Zook, H.~A., \& Jackson, A.~A. 1995, Icarus, 116, 186

\bibitem[{Love \& Brownlee(1993)}]{Love_1993}
Love, S.~G., \& Brownlee, D.~E. 1993, Science, 262, 550

\bibitem[{{Marsh} {et~al.}(2013){Marsh}, {Dalla}, {Kelly}, \&
  {Laitinen}}]{Marsh_etal_2013}
{Marsh}, M.~S., {Dalla}, S., {Kelly}, J., \& {Laitinen}, T. 2013, \apj, 774, 4,
  \dodoi{10.1088/0004-637X/774/1/4}

\bibitem[{McComas {et~al.}(2019)McComas, Rankin, Schwadron, \&
  Swaczyna}]{McComas_2019}
McComas, D.~J., Rankin, J.~S., Schwadron, N.~A., \& Swaczyna, P. 2019,
  Astrophys. J., 884, \dodoi{10.3847/1538-4357/ab441a}

\bibitem[{{Meyer}(1985)}]{Meyer_etal_1985}
{Meyer}, J.~P. 1985, \apjs, 57, 151, \dodoi{10.1086/191000}

\bibitem[{Miller \& Fields(2022)}]{miller2022supernova}
Miller, J.~A., \& Fields, B.~D. 2022, Astrophys. J., 934,
  \dodoi{10.3847/1538-4357/ac77f1}

\bibitem[{Moro-Mart{\'\i}n \& Malhotra(2002)}]{Moro-Martin_2002}
Moro-Mart{\'\i}n, A., \& Malhotra, R. 2002, Astron. J., 124, 2305

\bibitem[{Moro-Mart{\'\i}n \& Malhotra(2003)}]{Moro-Martin_2003}
---. 2003, Astrophys. J., 125, 2255

\bibitem[{M{\"u}ller {et~al.}(2008)M{\"u}ller, Frisch, Fields, \&
  Zank}]{Muller_2008}
M{\"u}ller, H.-R., Frisch, P.~C., Fields, B.~D., \& Zank, G.~P. 2008, in {From
  the Outer Heliosphere to the Local Bubble}, ed. J.~L. Linksy, V.~V.
  Izmodenov, E.~{M\"obius}, \& R.~{von Steiger} ({Springer New York})

\bibitem[{M{\"u}ller {et~al.}(2006)M{\"u}ller, Frisch, Florinski, \& {G. P.
  Zank}}]{Muller_2006}
M{\"u}ller, H.-R., Frisch, P.~C., Florinski, V., \& {G. P. Zank}. 2006,
  Astrophys. J., 647, 1491

\bibitem[{Nesvorn{\'y} {et~al.}(2010)Nesvorn{\'y}, Jenniskens, Levison, Bottke,
  Vokrouhlick{\'y}, \& Gounelle}]{Nesvorny_2010}
Nesvorn{\'y}, D., Jenniskens, P., Levison, H.~F., {et~al.} 2010, Astrophys. J.,
  713, 816

\bibitem[{Nesvorn{\'y} {et~al.}(2011)Nesvorn{\'y}, Vokrouhlick{\'y},
  Pokorn{\'y}, \& Janches}]{Nesvorny_2011}
Nesvorn{\'y}, D., Vokrouhlick{\'y}, D., Pokorn{\'y}, P., \& Janches, D. 2011,
  Astrophys. J., 743

\bibitem[{Opher {et~al.}(2024)Opher, Loeb, Zucker, Goodman, Konietzka, Worden,
  Economo, Miller, Alves, Grone, Kornbleuth, Peek, \&
  Foley}]{opher2024heliosphere}
Opher, M., Loeb, A., Zucker, C., {et~al.} 2024, Astrophys. J., 972,
  \dodoi{10.3847/1538-4357/ad596e}

\bibitem[{{Pei} {et~al.}(2006){Pei}, {Jokipii}, \& {Giacalone}}]{Pei_etal_2006}
{Pei}, C., {Jokipii}, J.~R., \& {Giacalone}, J. 2006, \apj, 641, 1222,
  \dodoi{10.1086/427161}

\bibitem[{Petit {et~al.}(2011)}]{Petit_2011}
Petit, J., {et~al.} 2011, Astron. J., 142

\bibitem[{Piquette {et~al.}(2019)Piquette, Poppe, Bernardoni, Szalay, James,
  Hor{\'a}nyi, Stern, Weaver, Spencer, \& Olkin}]{Piquette_2019}
Piquette, M., Poppe, A.~R., Bernardoni, E., {et~al.} 2019, Icarus, 321, 116

\bibitem[{Pokorn{\'y} {et~al.}(2018)Pokorn{\'y}, Sarantos, \&
  Janches}]{Pokorny_2018}
Pokorn{\'y}, P., Sarantos, M., \& Janches, D. 2018, Astrophys. J., 863,
  \dodoi{doi.org/10.3847/1538-4357/aad051}

\bibitem[{Pokorn{\'y} {et~al.}(2014)Pokorn{\'y}, Vokrouhlick{\'y},
  Nesvorn{\'y}, Campbell-Brown, \& Brown}]{Pokorny_2014}
Pokorn{\'y}, P., Vokrouhlick{\'y}, D., Nesvorn{\'y}, D., Campbell-Brown, M., \&
  Brown, P. 2014, Astrophys. J., 789

\bibitem[{Poppe(2016)}]{Poppe_2016}
Poppe, A.~R. 2016, Icarus, 264, 369

\bibitem[{Poppe {et~al.}(2023)Poppe, Szabo, Imata, Keller, \&
  Christoffersen}]{poppe2023seps}
Poppe, A.~R., Szabo, P.~S., Imata, E.~R., Keller, L.~P., \& Christoffersen, R.
  2023, Astrophys. J. Lett., 958, \dodoi{10.3847/2041-8213/ad0cf6}

\bibitem[{Poppe {et~al.}(2019)Poppe, Lisse, Piquette, Zemcov, Hor{\'a}nyi,
  James, Szalay, Bernardoni, \& Stern}]{Poppe_2019b}
Poppe, A.~R., Lisse, C.~M., Piquette, M., {et~al.} 2019, Astrophys. J. Lett.,
  881, \dodoi{https://doi.org/10.3847/2041-8213/ab322a}

\bibitem[{Press {et~al.}(2007)Press, Teukolsky, Vetterling, \&
  Flannery}]{Press_2007}
Press, W.~H., Teukolsky, S.~A., Vetterling, W.~T., \& Flannery, B.~P. 2007,
  {Numerical Recipes: The Art of Scientific Computing}, 3rd edn. ({32 Avenue of
  the Americas, New York, NY, 10013-2473, USA}: {Cambridge University Press})

\bibitem[{Quinn {et~al.}(2018)Quinn, Schwadron, {M\"obius}, Taut, \&
  Berger}]{Quinn_2018}
Quinn, P.~R., Schwadron, N.~A., {M\"obius}, E., Taut, A., \& Berger, L. 2018,
  Astrophys. J., 861, \dodoi{10.3847/1538-4357/aac6ca}

\bibitem[{Reames(1999)}]{Reames_1999}
Reames, D.~V. 1999, Space Sci. Rev., 90, 413

\bibitem[{Reames(2013)}]{Reames_2013}
---. 2013, Space Sci. Rev., 175, 53

\bibitem[{Reames(2019)}]{Reames_2019}
---. 2019, Atoms, 7, \dodoi{10.3390/atoms7040104}

\bibitem[{Reames(2021)}]{reames2021solar}
---. 2021, Solar energetic particles: a modern primer on understanding sources,
  acceleration and propagation (Springer Nature)

\bibitem[{Sandford(1986)}]{Sandford_1986}
Sandford, S.~A. 1986, Icarus, 68, 377, \dodoi{10.1016/0019-1035(86)90045-X}

\bibitem[{Schwadron {et~al.}(2002)Schwadron, Combi, Huebner, \&
  McComas}]{Schwadron_2002}
Schwadron, N.~A., Combi, M., Huebner, W., \& McComas, D.~J. 2002, Geophys. Res.
  Lett., 29, \dodoi{10.1029/2002GL015829}

\bibitem[{Schwadron {et~al.}(2000)Schwadron, Geiss, Fisk, Gloeckler, Zurbuchen,
  \& {von Steiger}}]{Schwadron_2000}
Schwadron, N.~A., Geiss, J., Fisk, L.~A., {et~al.} 2000, J. Geophys. Res., 105,
  7465, \dodoi{10.1029/1999JA000225}

\bibitem[{Schwadron \& Gloeckler(2007)}]{Schwadron_2007}
Schwadron, N.~A., \& Gloeckler, G. 2007, Space Sci. Rev., 130, 283,
  \dodoi{10.1007/s11214-007-9166-6}

\bibitem[{Stark \& Kuchner(2009)}]{Stark_2009}
Stark, C.~C., \& Kuchner, M.~J. 2009, Astrophys. J., 707, 543

\bibitem[{{Stone} \& {Cummings}(1997)}]{stone1997acrs}
{Stone}, E.~C., \& {Cummings}, A.~C. 1997, in International Cosmic Ray
  Conference, Vol.~2, International Cosmic Ray Conference, 289

\bibitem[{Szenes {et~al.}(2010)Szenes, Kov\'acs, P\'ecz, \&
  Skuratov}]{Szenes_2010}
Szenes, G., Kov\'acs, V.~K., P\'ecz, B., \& Skuratov, V. 2010, Astrophys. J.,
  708, 288, \dodoi{10.1088/0004-637X/708/1/288}

\bibitem[{Thiel {et~al.}(1991)Thiel, Bradley, \& Spohr}]{Thiel_1991}
Thiel, K., Bradley, J.~P., \& Spohr, R. 1991, Nucl. Tracks Radiat. Meas., 19,
  709, \dodoi{10.1016/1359-0189(91)90298-V}

\bibitem[{Tylka {et~al.}(2013)Tylka, Malandraki, Dorrian, Ko, Marsden, Ng, \&
  Tranquille}]{tylka2013sep}
Tylka, A.~J., Malandraki, O.~E., Dorrian, G., {et~al.} 2013, Solar Phys., 285,
  251, \dodoi{10.1007/s11207-012-0064-z}

\bibitem[{Verkhoglyadova {et~al.}(2012)Verkhoglyadova, Li, Ao, \&
  Zank}]{verkhoglyadova2012radial}
Verkhoglyadova, O.~P., Li, G., Ao, X., \& Zank, G.~P. 2012, Astrophys. J., 757,
  \dodoi{10.1088/0004-637X/757/1/75}

\bibitem[{Vitense {et~al.}(2012)Vitense, Krivov, Kobayashi, \&
  {L\"ohne}}]{Vitense_2012}
Vitense, C., Krivov, A.~V., Kobayashi, H., \& {L\"ohne}, T. 2012, Astron.
  Astrophys., 540

\end{thebibliography}



\end{document}